\documentstyle[prl,aps,twocolumn,epsfig,floats]{revtex}

\newcommand{\be}{\begin{equation}}
\newcommand{\ee}{\end{equation}}
\newcommand{\BE}{\begin{eqnarray}}
\newcommand{\EE}{\end{eqnarray}}
\newcommand{\BEn}{\begin{eqnarray*}}
\newcommand{\EEn}{\end{eqnarray*}}
\newcommand{\barr}{\begin{array}} 
\newcommand{\earr}{\end{array}}
\newcommand{\bit}{\begin{itemize}}      
\newcommand{\eit}{\end{itemize}}
\newcommand{\bfl}{\begin{flusleft}}
\newcommand{\efl}{\end{flusleft}}
\newcommand{\bfr}{\begin{flushright}}
\newcommand{\efr}{\end{flushright}}
\newcommand{\bfig}{\begin{figure}}
\newcommand{\efig}{\end{figure}}

\newcommand{\bc}{\begin{center}}
\newcommand{\ec}{\end{center}}

\newcommand{\ben}{\begin{enumerate}}    
\newcommand{\een}{\end{enumerate}}

\newcommand{\eps}{\varepsilon}
\newcommand{\de}{\partial}

\newcommand{\e}{{\mbox{e}}}
\newcommand{\erf}{{\mbox{Erf}}}

\begin{document}
\tighten
\draft
\twocolumn[\hsize\textwidth\columnwidth\hsize\csname@twocolumnfalse\endcsname

\title{Nonequilibrium Phase Transition in Non-Local and Nonlinear Diffusion 
Model}

\author{Fabio Cecconi$^{(1)}$, Jayanth R. Banavar$^{(2)}$ and 
        Amos Maritan$^{(1)}$}

\address{$(1)$ Intenational School for Advanced Studies (SISSA/ISAS), 
               INFM  and \\ ``The Abdus Salam'' 
               International Centre for Theoretical Physics \\ 
         $(2)$ Department of Physics and Center for Material Physics,
               104 Davey Laboratory, The Pennsylvania State University.}

\date{\today}
\maketitle

\begin{abstract}
We present the results of analytical and numerical studies of
a one-dimensional nonlocal and nonlinear
diffusion equation describing non-equilibrium processes ranging from 
aggregation phenomena to cooperation of individuals. 
We study a dynamical phase transition that is obtained
on tuning the initial conditions and demonstrate universality
and characterize the critical behavior.
The critical state is shown to be reached in a self-organized manner
on dynamically evolving the diffusion equation subjected
to a mirror symmmetry transformation.
\end{abstract}
]

The study of equilibrium phase transitions is now a
mature field \cite{PhaseT,Ma}
and, increasingly, attention is being paid to outstanding
problems in nonequilibrium statistical physics \cite{Zia}.  
Such problems are often
challenging  because of inherent non-linearities.
There are many examples of non-equilibrium phenomena
which are intrinsically non-local such as the growth of
thin films in the presence of shadowing \cite{Marsili} and the
sculpting of the drainage basin of river networks due
to erosional processes \cite{Banavar}.  A striking development
in the field of non-equilibrium statistical physics is the
development of the paradigm of self-organized criticality \cite{Bak},
entailing the competition between two dynamical processes
leading to a critical state without any fine tuning
of parameters.

The principal theme of this letter is the study of a simple one dimensional
nonlinear and nonlocal diffusion equation 
to elucidate the
nature of a non-equilibrium phase transition.
The equation is essentially one describing biased diffusion \cite{Chandra}
with the magnitude of the bias determined by the instantaneous
configuration of the random walkers undergoing diffusion.
Physically, the equation describes aggregation,
population dynamics, and represents a simple model
for cooperative behavior in game theory \cite{Axel}.
Strikingly, the behavior changes from that of a conventional
critical point (which requires tuning) to that of self-organized
criticality \cite{Bak} on considering the time evolution of a 
transformed equation obtained by a mirror symmetry transformation with $x$ 
replaced by $-x$.

Our basic equation for $P(x,t)$, the probability that the diffusing
particle is at position  $x$ at time $t$, is
\be
\frac{\de P}{\de t} = - v(t)  \frac{\de P}{\de x} + 
\frac{1}{4}\frac{\de^2 P}{\de x^2}   
\label{eq:Pcont}
\ee
where both the nonlinearity as well as the nonlocality
are introduced in the bias velocity $v$ defined by
\be
v(t) = \int_0^{\infty} dx P(x,t) - 1/2 \;. 
\label{eq:veldef}
\ee
On setting $v$ equal to zero in Eq.~(\ref{eq:Pcont}), one recovers the 
standard unbiased diffusion equation \cite{Chandra,BarNin,MSW}, 
whereas one obtains simple biased diffusion,
when $v$ is a constant. In our equation, $v$ is a measure of the 
imbalance between the population of walkers in the right and left and  
the drift bias promotes further aggregation.  

Eq.~(\ref{eq:Pcont}) describes the temporal evolution of the distribution 
function $P(x,t)$ and leads to one of two outcomes in the large time limit.
Depending on the initial distribution, one ends up
with the bias to the right or to the left winning
so that $P$ becomes 1 either at $+\infty$
or at $-\infty$.  Our focus is on the non-equilibrium transition between 
these limiting behaviors.
Note that there is a set of initial conditions (of measure zero) that 
correspond to the dynamical critical point -- we will demonstrate that the 
critical behavior is universal. 
 
Let us define a new variable $w(t)=\int_0^t d\tau v(\tau)$, and introduce 
$y(t)= x-w(t)$  so that Eq.~(\ref{eq:Pcont}) is cast in the form of
a standard diffusion equation,
\be
\frac{\de P}{\de t} = 
\frac{1}{4}\frac{\de^2 P}{\de y^2}\; .  
\label{eq:diff}
\ee
For simplicity, let us first consider an initial Gaussian distribution of 
$P(x,t)$ centered around $x_0$ and with variance $\sigma_0$.
The solution of Eq.~(\ref{eq:diff})
is then given by 
\be
P(x,t) = N \exp\bigg\{-\frac{[x-x_0 - w(t)]^2}{t + 2\sigma_0^2}\bigg\}\;.
\label{eq:Pxt}
\ee
where $N=1/\sqrt{\pi(t+2\sigma_0^2)}$ is the normalization constant.

With $\sigma_0 = 0$, $P(x,t)$ is also the fundamental solution \cite{Risken}, 
which will be used to obtain the distribution at time $t$, starting from more 
general initial conditions.
Expression~(\ref{eq:Pxt}) is only a formal solution because 
$w(t)$ is itself a function  of $P(x,t)$.
The transition is between two phases corresponding to aggregation 
on the right or on the left and therefore one would expect that the 
critical point would correspond to a situation with no bias 
(i.e. $x_0 = 0$).
In this case, the distribution is symmetric with respect to the origin 
at all times, and there is nothing to choose between left and right, thus an 
unbiased behaviour ensues. 
In order to probe the behaviour in the vicinity of this critical 
point, one could start with an initial distribution with a tiny value of 
$x_0$ (small bias) and watch how the system evolves.   
Combining Eqs.~(\ref{eq:Pcont}) and~(\ref{eq:veldef}) and noting that 
$\dot{w} = v$, we find
\be
\dot{w} = \frac{1}{2}\erf \bigg\{\frac{w(t)+x_0}{\sqrt{t+2\sigma_0^2}}\bigg\}   
\label{eq:wdot}
\ee
At criticality ($x_0=0$), the velocity is zero at all times and $w$ vanishes 
too. In the critical regime, one expects that $w(t) \ll 1$, for all $t$ less 
than a crossover time $T_r$, so that one may linearize Eq.~(\ref{eq:wdot}) 
to find that
\be
\dot{w} = \frac{1}{\sqrt{\pi}}\frac{w + x_0}{\sqrt{t+2\sigma_0^2}}.
\ee
This  equation can be easily solved and yields
\be
w(t) = x_0 \bigg\{\e^{2(\sqrt{t+2\sigma_0^2} - 
\sqrt{2\sigma_0^2})/\sqrt{\pi}} - 1 \bigg\}    .
\label{eq:wlin}
\ee
One may define a characteristic transient time, $T_r$, spent in the critical
region during which the linearization approximation holds.
From~(\ref{eq:wlin}), one finds 
that $T_r$ diverges as $x_0\to 0$  as 
\be
T_r \sim \ln^2|x_0|. 
\label{eq:Tr}
\ee 
In the critical region, the characteristic length-scale $\xi$ is 
expected to follow the diffusion law, and therefore   
\be
\xi \sim \sqrt{T_r} \sim |\ln|x_0||\;,
\ee 
as we will verify  numerically  later on. 

Alternatively, in the initial condition,
one may introduce a bias by fixing $x_0=0$ and instead 
letting $P(\infty,0) = \Phi_0$ as an effective asymmetric boundary condition. 
With the same procedure used to derive Eq.~(\ref{eq:wdot}) we have
\be
\dot{w} = \frac{\Phi_0}{2} + 
\frac{1-\Phi_0}{2}\erf \bigg\{\frac{w(t)}{\sqrt{t+2\sigma_0^2}}\bigg\}\;.
\label{eq:wdot1}
\ee
$\Phi_0 = 0$ corresponds to the critical point, and by linearizing 
Eq.~(\ref{eq:wdot1}) for small $\Phi_0$, it is straightforward to show that 
one again obtains $T_r \sim \ln^2(\Phi_0)$, with $\xi \sim \sqrt{T_r}$.

We have verified that the same critical behaviour holds for other families 
of initial conditions. For example, when  
\be
P(x,0) = (1/2 - \eps)\; \delta(x-1) + (1/2 + \eps)\; \delta(x+1)   ,
\label{eq:P0eps}
\ee
we find that
\be
T_r \sim \ln^2|\eps| 
\label{eq:Tofeps}
\ee 
and $\xi \sim |\ln |\eps||$. 
Note that $\eps$ is now a measure of the deviation from the critical point 
and $\eps=0$  correponds to the unbiased, zero drift situation.  

We now turn to the results of numerical experiments on a 1-dimensional 
lattice which are useful for validating our analytic predictions and for 
probing the nonlinear regime. 
The discrete version of Eq.~(\ref{eq:Pcont}), used in our simulations, reads 
\be
P_x(t+1) = \frac{P_x(t)}{2} + \frac{\Phi(t)}{2} P_{x-1}(t) + 
\frac{1-\Phi(t)}{2} P_{x+1}(t)
\label{eq:Pdis}
\ee
with $\Phi(t) = \sum_{x\geq 0} P_x(t)$. The velocity is thus given by 
$v(t) = \Phi(t) - 1/2$. 
This equation was proposed by Nowak and Sigmund 
\cite{NowSig98}
as a simplified model  of the evolution of indirect 
reciprocity by image scoring. Indirect reciprocity is determined by 
reputation and status and is characterized by each individual having an 
image score. 
A potential donor and recipient of an altruistic act have an interaction  
in which the donor helps the recipient provided the recipient's image score
is positive. 
Such an altruistic act increases the image score of the donor by $1$
(the selfish act would have decreased it by $1$) and the image score of the 
recipients is unchanged.  Eq.~(\ref{eq:Pdis}) is the governing equation 
for the time evolution of $P_x$, the fraction of players with image score $x$.
The two phases that we have considered correspond to cooperation and defection
and our finding is that not much time is spent agonizing over which phase 
to select even in the vicinity of the critical point. Indeed,
the time scale to decide on one of the two different  phases  only diverges 
logarithmically as one approaches the critical point.
\bfig
\bc
\epsfig{figure=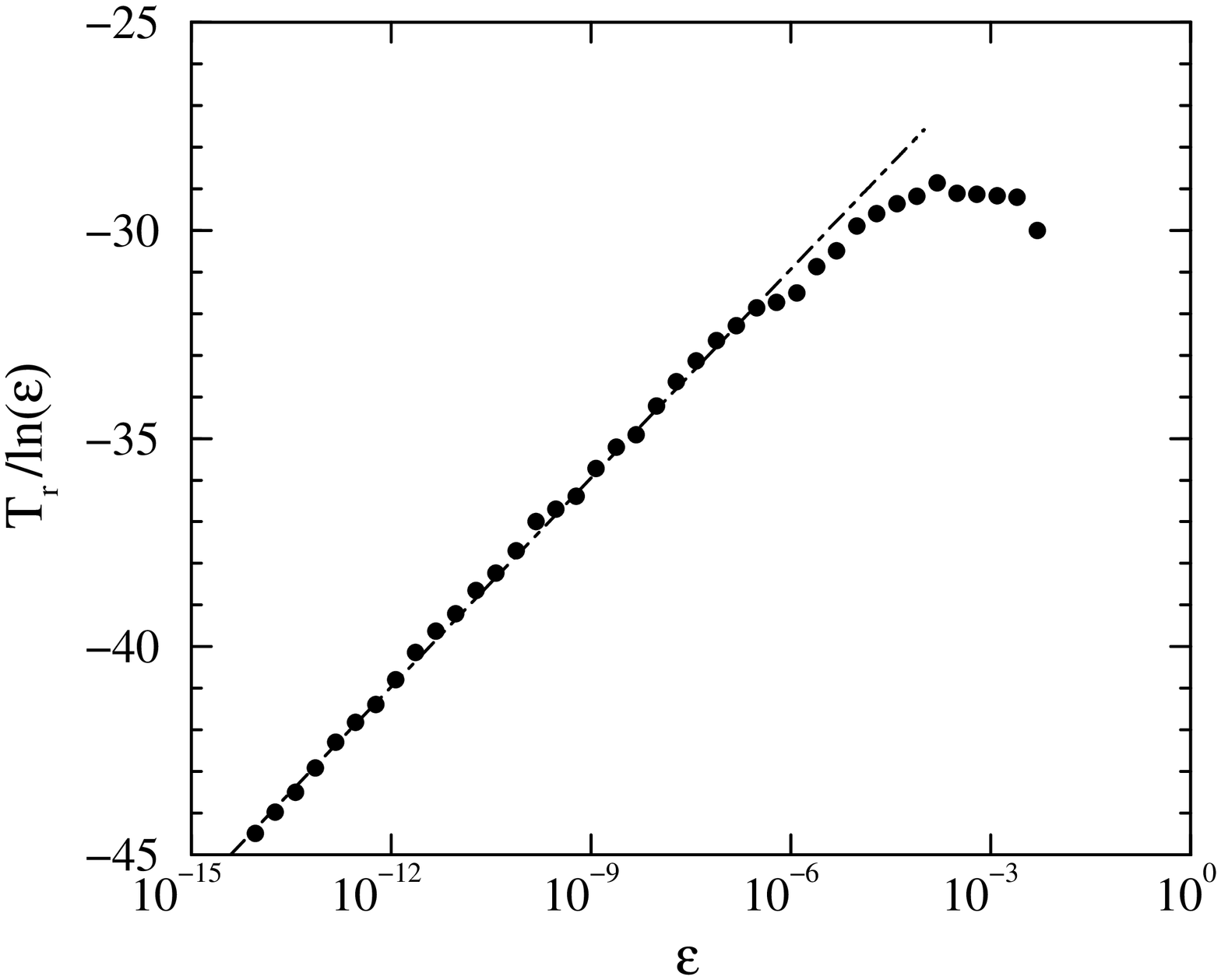,height=1.1\columnwidth,angle=270}
\ec
\caption{Behaviour of transient time $T_r$ (at which $v(t)$ first becomes
equal to $1/2$),  
for the initial distribution (\ref{eq:P0eps}) for different
$\eps$ values. In the plot, the square logarithmic divergence corresponds
to linear behaviour.}
\label{fig:scaleps}
\efig 

Fig.~\ref{fig:scaleps} shows the divergence of $T_r$ as $\eps \to 0$ for 
the initial distributions given in Eq.~(\ref{eq:P0eps}) and for a lattice 
of $6$x$10^5$ sites. Numerical results are in excellent
accord with our theoretical predictions (see Eq.~(\ref{eq:Tofeps}))

Let us consider now, an interesting generalization of the initial conditions 
that we studied analytically: 
$P(\infty,0) = \Phi_0$ and $P(-z,0) = 1-\Phi_0$ and $P=0$ at all other 
locations, initially.
The critical value of $\Phi_0$ increases monotonically to a 
non-zero value $\Phi_c(z)$ for positive $z$ with $\Phi_c(z\to \infty)=1/2$.
The smallest value of $\Phi_c(z)$ on a discrete lattice occurs when $z=1$, 
which is the case we focus on. 
$\Phi_c \equiv \Phi_c(z=1)$ is found to be 0.261970531164..., a result that 
was noted earlier by Nowak and Sigmund \cite{NowSig98}.   

Fig.~\ref{fig:veloc} shows the behaviour of the bias velocity as a 
function of time as $\Phi_0$ approaches $\Phi_c$ from above and from below. 
\bfig
\bc
\epsfig{figure=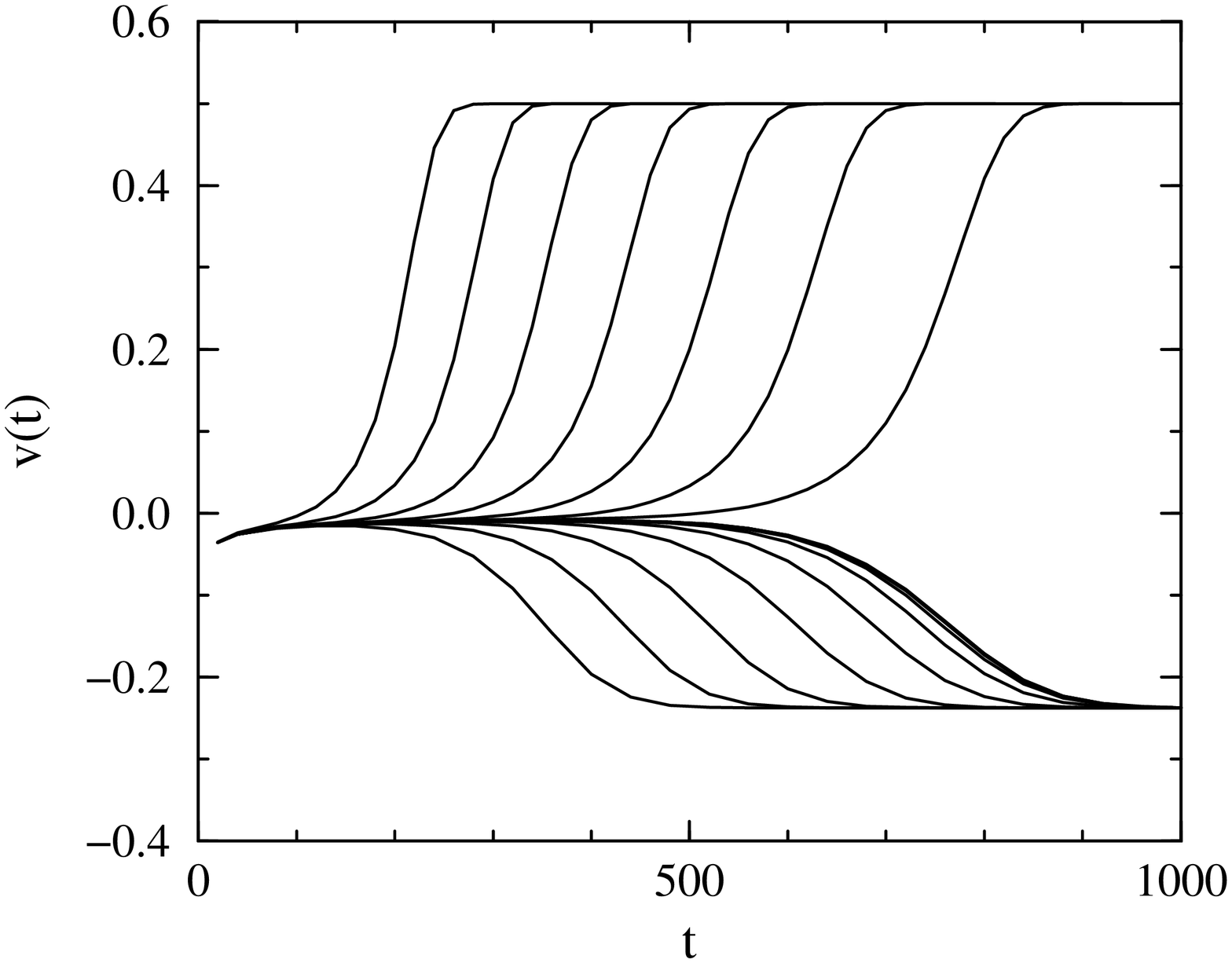,height=1.1\columnwidth,angle=270}
\ec
\caption{Temporal behaviour of the bias velocity $v(t)$ for various
values of the  parameter 
$\Phi_0$  around the critical value $\Phi_c$. 
The upper curves correspond to values of $\Phi_0$ approaching $\Phi_c$ 
from above. The lower curves are obtained when $\Phi_0$ 
tends to $\Phi_c$ from below.}
\label{fig:veloc}
\efig 
$T_r$ can be identified as the time after which $v(t)$ becomes equal to 
its asymptotic values of either $1/2$ and $\Phi_0 -1/2$  and its scaling 
behaviour is shown in Fig.~\ref{fig:scale}.
\bfig[h]
\bc
\epsfig{figure=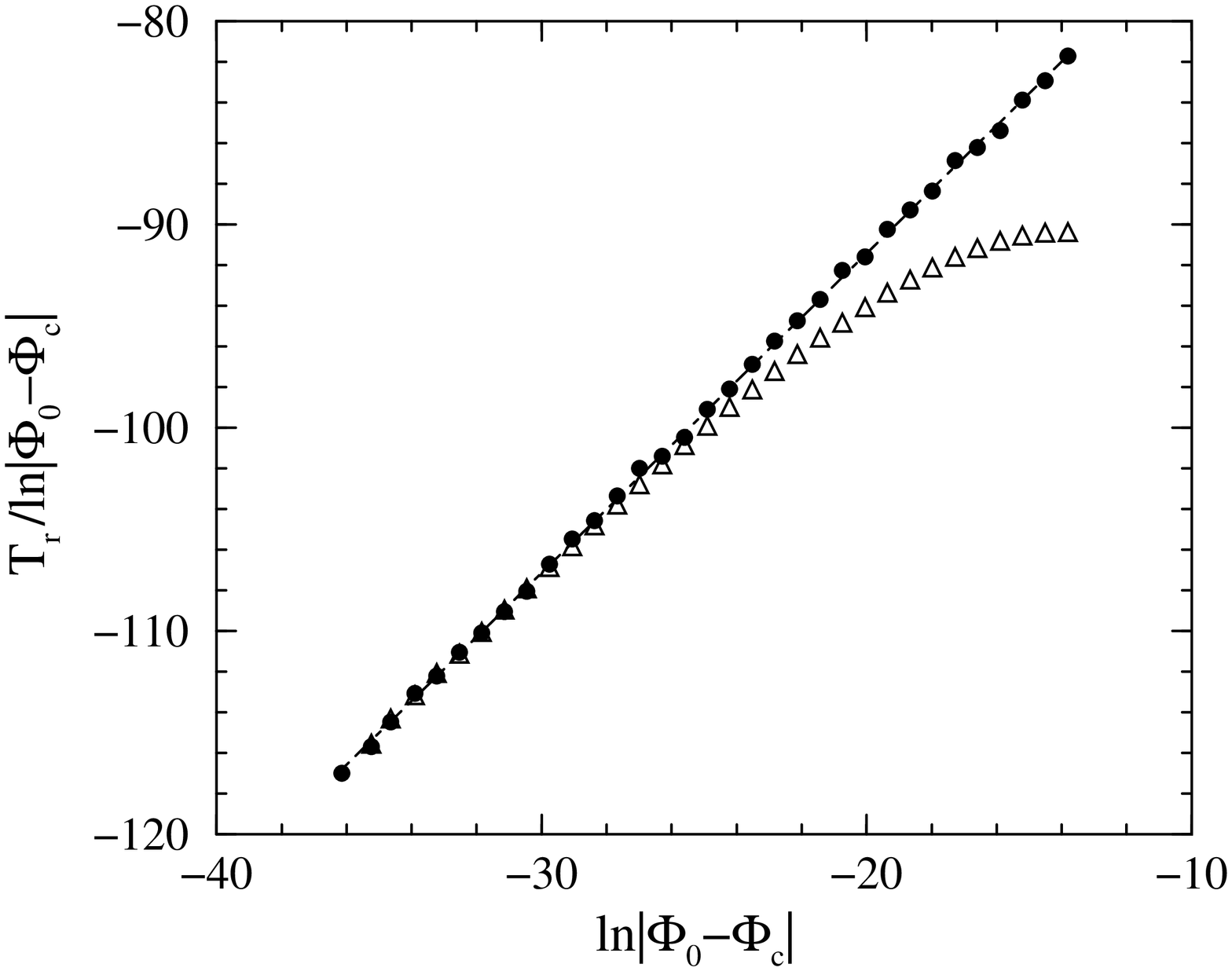,height=1.1\columnwidth,angle=270}
\ec
\caption{Dependence of the transient time $T_r$ on $(\Phi_c-\Phi_0)$. $T_r$ 
is defined as the time needed for $v(t)$ to approach the value $1/2$ when 
$\Phi_0>\Phi_c$ and $\Phi_0-1/2$ for $\Phi_0 <\Phi_c$.
The bullets show the behavior when $\Phi_0<\Phi_c$ while the
triangles correspond
to  $\Phi_0 > \Phi_c$.}
\label{fig:scale}
\efig 
The average location of the random walkers (excluding the 
number fixed at $x=\infty$) behaves with time (in the vicinity of the 
fixed point) as 
\be
\langle x(t) \rangle \sim  \sqrt{t+2\sigma^2_0}\;.  
\label{eq:xav}
\ee

In order to derive this result, we note that 
the velocity increases very slowly in the linear critical regime
and can be approximately considered constant (this is consistent with the 
assumption that, for time $t\ll T_r$, $w$ is much smaller than $1$).
The derivative of the velocity given by
Eq.~(\ref{eq:veldef}) is then vanishing, and using 
Eq.~(\ref{eq:Pcont}) to eliminate $\dot{P}$ we obtain, after 
integrating over $x$, the expression 
\be
 4 v P(0,t) - \de_x P(0,t) = 0\;.
\ee
Using the formal solution~(\ref{eq:Pxt}), we obtain 
$$
w(t) \sim \sqrt{t + 2\sigma_0^2}
$$
The result~(\ref{eq:xav}) then follows on noting that the average
position of the walkers is given by $w(t)+x_0$.

\bfig[h]
\bc
\epsfig{figure=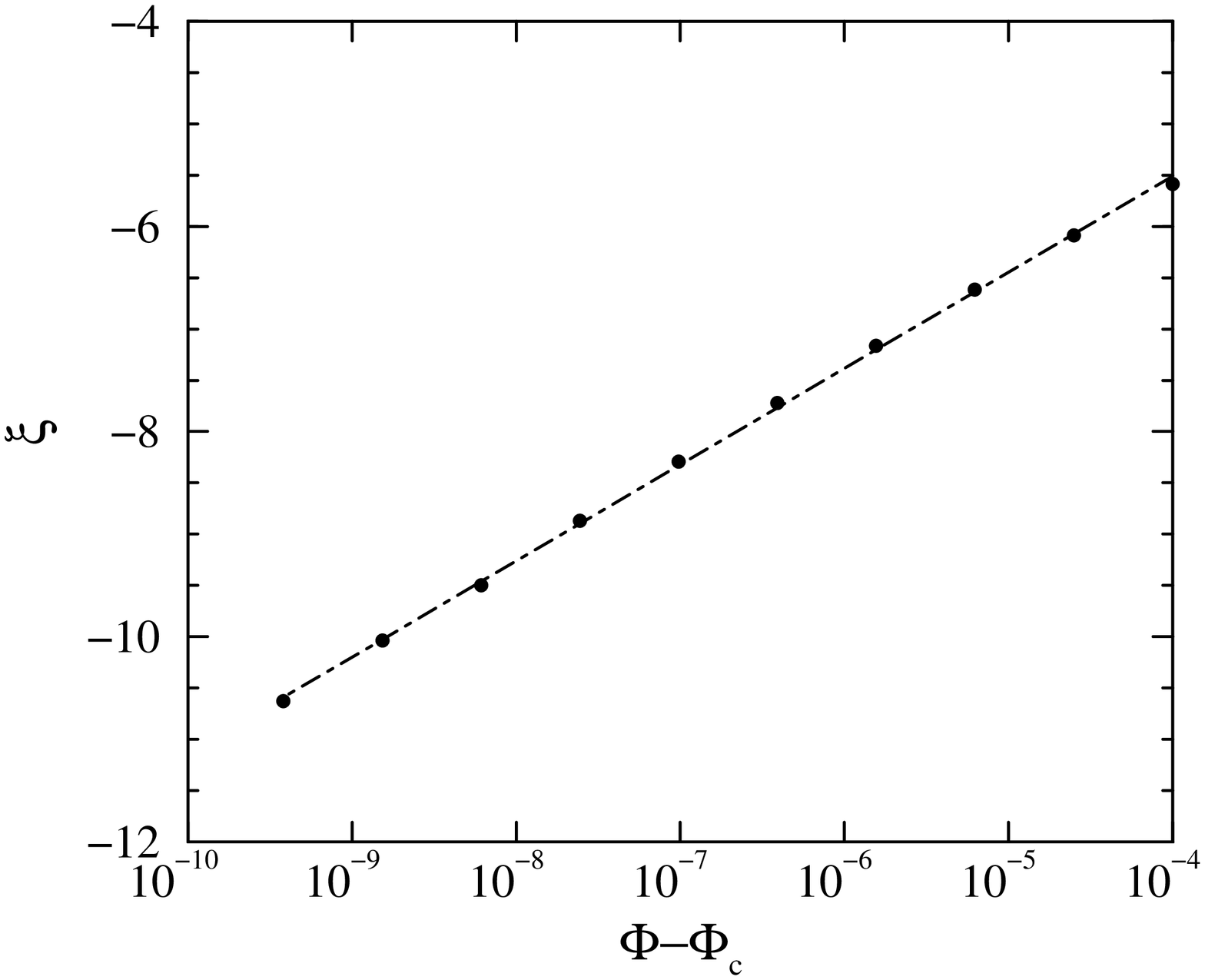,height=1.1\columnwidth,angle=270}
\ec
\caption{Plot of the characteristic length $\xi$, as a function 
of the deviation from criticality $\Phi_c-\Phi_0$. This numerical
result confirms that, in the critical regime, $\xi$ 
diverges logarithmically as the critical point is approached.}
\label{fig:length}
\efig 

This critical regime behaviour of $\langle x(t) \rangle$ crosses
over to a linear temporal behavior when the bias 
reaches a sufficient strength. There is indeed a sharp onset of the linear
behaviour at a value of $\langle x \rangle$, which one may identify with 
$\xi$. The scaling behaviour of $\xi$ is shown in Fig.~\ref{fig:length}.

We now turn to a simple mechanism for obtaining self-organized 
critical behavior in our model.
In Eq.~(\ref{eq:Pcont}) the transformation 
$x \to -x$ is equivalent to a change of 
the sign of the bias velocity. 
In this situation, the system spontaneously 
organizes in such a way that the 
aggregation of walkers is disfavoured. 
As a consequence, the asymptotic 
distribution becomes symmetric (characterized by $v=0$)
and this corresponds to a self-tuning to the 
critical state, a behaviour typical of self-organized 
criticality \cite{Bak}.

\bfig
\bc
\epsfig{figure=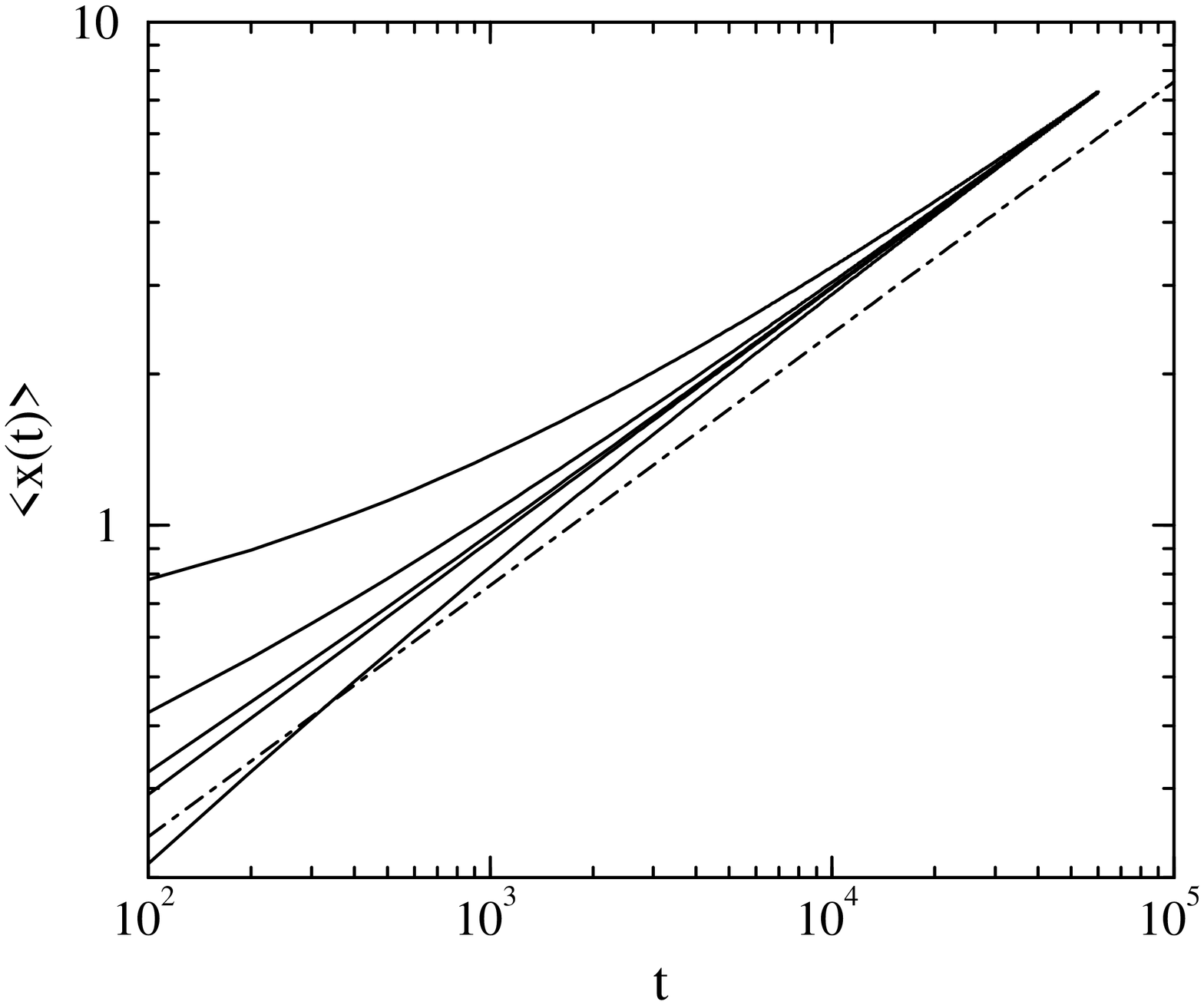,height=1.1\columnwidth,angle=270}
\ec
\caption{Temporal behavior of the average location of walkers 
(excluding the fraction $\Phi_0$ placed at $x=\infty$)
when the sign of the velocity in Eq.~(\ref{eq:Pcont}) 
is changed. The plot is obtained for the initial conditions 
$P(x=-1) = 1-\Phi_0$ for four values of $\Phi_0$ randomly choosen
in the interval $[0,1/2]$.   
The dot-dashed line indicates the predicted $\sqrt{t}$-behaviour.}
\label{fig:mvavx}
\efig 

In this asymptotic regime, the scaling $\langle x(t) 
\rangle \sim \sqrt{t}$ derived in (\ref{eq:xav}) still ought to
hold and is confirmed by 
simulations performed on the discretized diffusion 
equation (see Fig.~\ref{fig:mvavx}).

In this letter, we have introduced and studied a diffusion equation with
nonlinear and nonlocal features.
In this model, spontaneous fluctuations in the population of walkers are
able to drive the entire population towards one of the two
boundaries located at $(x=\pm \infty)$.
This mechanism is of interest as the basis for the development of more
realistic models of self-aggregation and self-organization in cooperative
states of populations of interacting individuals.
A mirror symmetry transformation 
applied to the equation reveals a dynamical evolution
corresponding to generic behavior associated with self-organized
criticality.

This work was supported  by INFN, NASA, NATO and The Donors of the 
Petroleum Research Fund administered by the American Chemical Society. 


\end{document}